\documentclass[fleqn,usenatbib]{mnras}

\usepackage{newtxtext,newtxmath}

\usepackage[T1]{fontenc}
\usepackage{ae,aecompl}

\usepackage{graphicx}	
\usepackage{amsmath}	
\usepackage{amssymb}	

\newcommand{\Ni}{\ensuremath{^{56}\mathrm{Ni}}}

\newcommand{\Msun}{\ensuremath{\mathrm{M}_\odot}}

\newcommand{\kmps}{\ensuremath{\mathrm{km~s^{-1}}}}

\newcommand{\Msunpyr}{\ensuremath{\Msun~\mathrm{yr^{-1}}}}

\title[CSM around SNe Ia with He star donors]{
Circumstellar properties of Type~Ia supernovae from the helium star donor channel
}

\author[T. J. Moriya et al.]{
Takashi J. Moriya,$^{1}$\thanks{E-mail: takashi.moriya@nao.ac.jp (TJM)}
Dongdong Liu,$^{2,3,4,5}$
Bo Wang$^{2,3,4,5}$
and Zheng-Wei Liu$^{2,3,4,5}$
\\
$^{1}$Division of Science, National Astronomical Observatory of Japan, National Institutes of Natural Sciences, 2-21-1 Osawa, Mitaka,\\ Tokyo 181-8588, Japan\\
$^{2}$Yunnan Observatories, Chinese Academy of Sciences, Kunming 650216, China \\
$^{3}$Key Laboratory for the Structure and Evolution of Celestial Objects, Chinese Academy of Sciences, Kunming 650216, China\\
$^{4}$University of Chinese Academy of Sciences, Beijing 100049, China\\
$^{5}$Centre for Astronomical Mega-Science, Chinese Academy of Sciences, Beijing 100012, China
}

\date{Accepted 2019 July 06. Received 2019 July 06; in original form 2019 April 09}

\pubyear{2019}

\begin{document}
\label{firstpage}
\pagerange{\pageref{firstpage}--\pageref{lastpage}}
\maketitle

\begin{abstract}
We investigate predicted circumstellar properties of Type~Ia supernova progenitor systems with non-degenerate helium star donors.
It has been suggested that systems consisting of a carbon+oxygen white dwarf and a helium star can lead to Type~Ia supernova explosions.
Binary evolution calculations for the helium star donor channel predict that such a progenitor system is in either a stable helium-shell burning phase or a weak helium-shell flash phase at the time of the Type~Ia supernova explosion. 
By taking the binary evolution models from our previous study, we show that a large fraction of the progenitor systems with a helium star donor have low enough density to explain the current non-detection of radio emission from Type~Ia supernovae. Most of the progenitor systems in the weak helium-shell flash phase at the time of the Type~Ia supernova explosions, which may dominate the prompt (short delay time) Type~Ia supernova population, have both low circumstellar density and a faint helium star donor to account for the non-detection of radio emission and a pre-explosion companion star in SN~2011fe and SN~2014J. 
We also find some progenitor systems that are consistent with the properties of the companion star candidate identified at the explosion location of Type~Iax SN~2012Z. 
\end{abstract}




\section{Introduction}
Type~Ia supernovae (SNe) are suggested to be thermonuclear explosions of C+O white dwarfs (WDs, e.g., \citealt{hoyle1960fowler}), which is recently confirmed observationally \citep{nugent2011iawd,bloom2012sn11fepro}. 
Despite of their uniformity that led to the discovery of the accelerating expansion of the Universe \citep{riess1998,perlmutter1999}, there are diverse theories for their progenitors and explosion mechanisms (see \citealt{maoz2014iareview,maeda2016iareview,livio2018iareview,wang2018sniarev} for recent reviews). 

One big question in SN~Ia progenitors is the nature of their companion stars. 
The canonical model of SNe~Ia requires C+O WDs to grow their mass near the Chandrasekhar limit to trigger thermonuclear explosions \citep[e.g.,][]{thielemann2004iarev}. 
The nature of the donor stars is, however, largely debated. One possible donor star is a non-degenerate star (the single-degenerate (SD) model; e.g., \citealt{nomoto1982sd,whelan1973sd}). The non-degenerate star can transfer its mass to the WD so that the WD can burn the transferred mass on its surface to grow its mass. Another possibility is that the companion star is also a WD (the double-degenerate (DD) model; e.g., \citealt{webbink1984dd,iben1984dd}). The companion WD itself could be a donor star enabling a stable burning \citep[e.g.,][]{bildsten2007pia} or two WDs merge after losing their orbital energy through gravitational waves. The merger itself may trigger a SN~Ia explosion \citep[e.g.,][]{pakmor2012vioia} or a disrupted WD during the merger may be accreted to the survived WD to make it grow near the Chandrasekhar limit \citep[e.g.,][]{dan2014wdmerger,shen2015wdmerger,sato2016wdmerger,liu2018iadd,mori2019wdmerger}.
However, the later case is usually considered to end up with accretion-induced collapse (AIC) of WDs rather than SN~Ia explosions (e.g., \citealt{nomoto1985off,saio1985off}, but see also \citealt{yoon2007mergedwd}).

Many observational attempts have been made to constrain the companion stars. One method often used is radio and X-ray observations. The SD model is predicted to have much denser circumstellar media (CSM) than the DD model because of the mass transfer required to grow WDs. The interaction between SN ejecta and a CSM results in radio and X-ray emission through which we can constrain CSM properties around SNe \citep[e.g.,][]{chevalier1998ssa,chevalier2006csmrad,maeda2012shock}. Especially, many attempts have been made to observe SNe~Ia in radio but no radio signals have been detected from SNe~Ia (\citealt{chomiuk2014sn14jradio,chomiuk2016ialimits,horesh2012sn11feradiox,perez-torres2014sn14jradio} and references therein). These radio observations exclude most of CSM properties predicted by the SD model with H-rich star donors and they are proposed to favor the DD model, although the SD H-rich star donor model also has a way to make a low CSM density environment if rotation of accreting WDs and their spin-down time are taken into account \citep[e.g.,][]{justham2011}.

Although observations have been mainly compared with the SD model with H-rich star donors, it has been suggested that accretion from non-degenerate He donor stars can also make the accompanying WDs to reach near the Chandrasekhar limit
\citep[e.g.,][]{iben1994hedon,yoon2003hedon,wang2009heprog,brooks2016heacc,2016ApJ...819..168H,wong2019schwab,neuntefuel2019}. Especially, population synthesis models generally predict that SNe~Ia from the SD He star donor channel are likely dominant in the prompt (short delay time) population of SNe~Ia \citep[e.g.,][]{wang2009hepop,ruiter2009iapop,claeys2014iapop,liu2015sdiax}.
WDs with He star donors have also been related to peculiar kinds of thermonuclear SNe such as SN~2002cx-like (a.k.a. Type~Iax) SNe (\citealt{foley2013iax} and references therein) and a possible He star donor has been identified in the pre-explosion image of a SN Iax 2012Z \citep{mccully2014sn12Z,stritzinger2015sn12Z,yamanaka2015sn12Z}.

In the previous studies investigating CSM properties of SNe~Ia, CSM properties expected from the SD He donor channel have not been considered much. Here, we investigate its CSM properties based on a grid of He star donor SN~Ia progenitor models computed by \citet{wang2009heprog}. We compare our results with the previous constraints on the SN~Ia CSM properties and show that most of the He star donor progenitors are hard to be excluded even with the current deepest radio observations.

The rest of this paper is organized as follows. We first introduce the binary evolution model of \citet{wang2009heprog} which we use in this study in Section~\ref{sec:evmodel}. We estimate the CSM properties around He star donor systems at the time of SN~Ia explosions in Section~\ref{sec:csm}. We compare the results with observations and have general discussion on the He star donor model for SNe~Ia in Section~\ref{sec:discussion}. We conclude this paper in Section~\ref{sec:conclusions}.

\begin{figure*}
	\includegraphics[width=2\columnwidth]{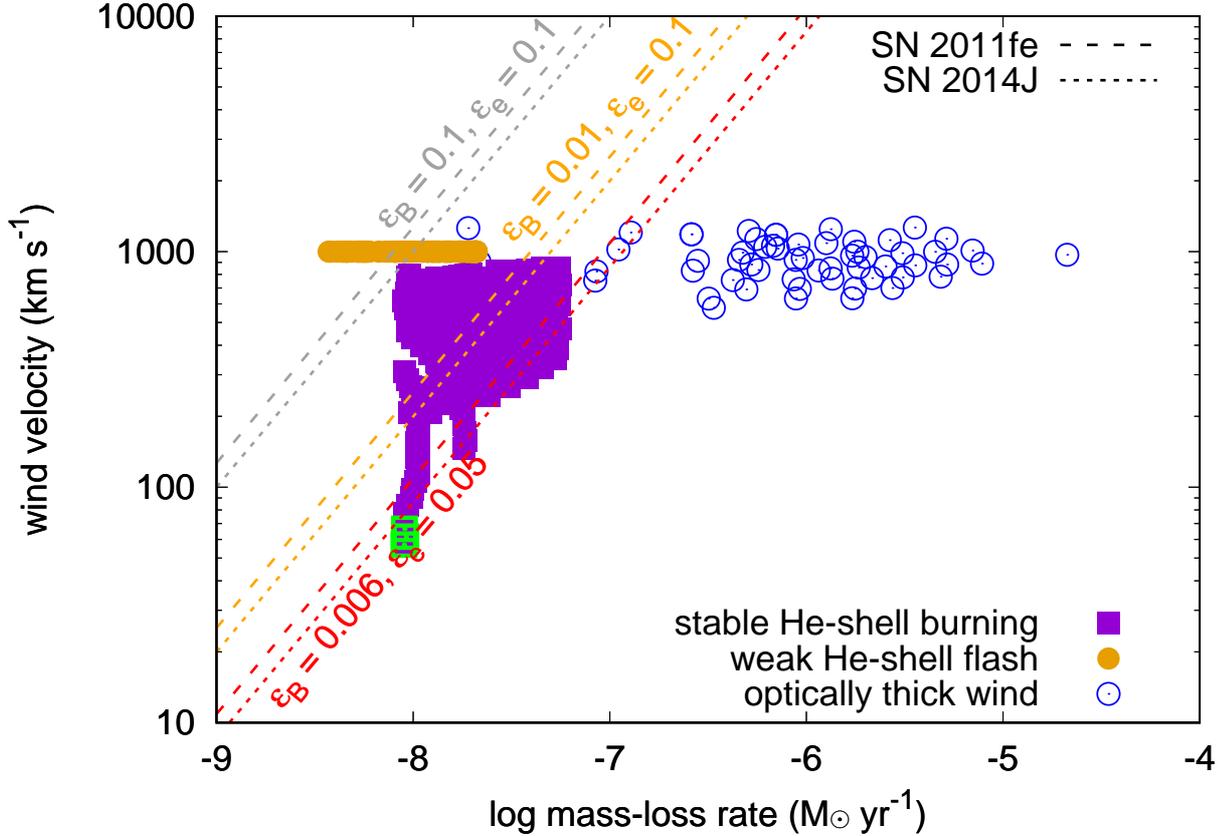}
    \caption{
    Mass-loss rates and wind velocities of SN~Ia progenitors with He star donors. The burning phase of each system is indicated by the symbols. The observational constraints on the CSM properties from SN~2011fe and SN~2014J are shown with lines with three different microphysics ($\varepsilon_B$ and $\varepsilon_e$) assumptions. The right side region of the lines is excluded by the radio observations. The models surrounded by green are those consistent with the companion star of SN~2012Z as shown in Fig.~\ref{fig:hrdbc}.
    }
    \label{fig:allch}
\end{figure*}

\section{Binary evolution model}\label{sec:evmodel}
We adopt the binary evolution models of \citet{wang2009heprog} in this study.
We refer to \citet{wang2009heprog} for the full details of the assumptions in the stellar and binary evolution calculations and we only briefly summarize them here. 
\citet{wang2009heprog} followed the binary evolution of about 2600 C+O WD + He star systems 
to identify the systems leading to SNe~Ia.
A SN~Ia explosion is assumed to occur when the mass of the C+O WD in a system reaches 1.378~\Msun. The Eggleton's stellar evolution code \citep{eggleton1971,eggleton1972,eggleton1973,han1994,pols1995,pols1998} is used to compute the stellar evolution in binary systems. Briefly, the Roche-lobe overflow is treated by the method described in \citet{han2000}. Mixing length theory is adopted with the Schwarzschild 
criteria with the mixing length parameter of 2.0. No overshooting is considered. The He stars are originally composed of He abundance of $Y=0.98$ with the solar metallicity ($Z=0.02$).
The structure of WDs is not solved in the binary evolution calculation and the WD mass growth rate is set by the mass transfer rate.
WDs about to explode as SNe~Ia from the He star donor channel are expected to be in one of the following three burning phases: (1) the optically thick wind phase, (2) the stable He-shell burning phase, and (3) the weak He-shell flash phase.

The first important physical property determining the mass growth rate is the maximum He accretion rate that a WD can burn on the surface. It is $\dot{M}_\mathrm{max}\simeq 7.2\times 10^{-6}(M_\mathrm{WD}/\Msun-0.6)~\Msunpyr$, where $M_\mathrm{WD}$ is the accreting WD mass \citep{nomoto1982heacc}. When the mass transfer rate becomes larger than this critical rate, the WD cannot burn all the transferred He and a He envelope 
is formed on top of the WD. This phase is called ``optically thick wind phase'' and we assume the extended He envelope results in the so-called ``optically thick wind'' \citep[e.g.,][]{kato1994otwind,hachisu1996otwind}. The WD mass grows with $\dot{M}_\mathrm{max}$ in this phase.

If the He accretion rate onto the WD is less than $\dot{M}_\mathrm{max}$ but above the minimum stable He burning accretion rate ($\dot{M}_\mathrm{st}$, see \citealt{kato2004mgrate}), all the accreted He is burned at the WD surface and the WD mass grows steadily with the mass transfer rate. 
When the He accretion rate goes below $\dot{M}_\mathrm{st}$, the He shell flash occurs at the WD surface. If the accretion rate is below $\dot{M}_\mathrm{st}$ but above the weak He-shell flash accretion limit ($4.0\times 10^{-8}~\Msunpyr$, \citealt{woosley1986weakhe}), the He flash is ``weak'' and a part of the shell is ejected from the system. The WD mass can still grow in this phase with the rate estimated by \citet{kato2004mgrate}. When the mass transfer rate becomes lower than the weak He-shell flash accretion limit, the He-shell flash becomes violent enough to prevent the WD mass growth. The WD mass does not increase in this case. 

Although \citet{wang2009heprog} assumed that the WDs at the optically thick wind phase explode as SNe~Ia, \citet{wang2017heaic} later found that O+Ne+Mg WDs are usually formed during the optically thick wind phase and they do not explode as SNe~Ia. These O+Ne+Mg WDs end up with AIC (e.g., \citealt{nomoto1991aic}) rather than SN~Ia explosions. Because such a system leading to AIC is considered to explode as SNe~Ia in the original models in \citet{wang2009heprog}, we also show their CSM properties for reference in this study.

\section{Circumstellar properties}\label{sec:csm}
The two essential parameters in estimating the CSM property of each system are its mass-loss rate and wind velocity. In this section, we present how we estimate the mass-loss rate and wind velocity from each system from the binary evolution model introduced in the previous section. The way we use to estimate the mass-loss rates and wind velocities depends on in which phase the system is at the time of the SN~Ia explosions.

\subsection{Stable He-shell burning phase}\label{sec:stablehe}
If the system is in the stable He-shell burning phase, the accreted mass onto the WD from the He donor is all steadily burned on the surface of the WD. No mass loss is expected from the system in our binary evolution calculations. 
In reality, however, a small fraction of transferred mass is likely to escape from the system from the outer Lagrangian point \citep[e.g.,][]{huang1996yu,deufel1999olvel}. We assume that 1\% of the transferred mass will be lost through the outer Lagrangian point \citep{huang1996yu}. 
We note that mass loss from the outer Lagrangian point leads to orbital shrinking of binary systems and, therefore, may result in a different evolution. However, we do not take such a difference into account in this work, partly because of the uncertain fraction of mass lost through the outer Lagrangian point. Especially, such a mass loss could change the initial binary parameter range of binary systems leading to SN~Ia explosions during the stable He-shell burning phase. Under these assumptions, 
the mass-loss rates are estimated to be between $\sim 10^{-8}~\Msunpyr$ and $\sim 10^{-7}~\Msunpyr$ (Fig.~\ref{fig:allch}), because
the mass transfer rates of the systems in this phase at the time of the SN~Ia explosions are between $\sim 10^{-6}~\Msunpyr$ and $\sim 10^{-5}~\Msunpyr$ in our binary evolution models.

The wind velocity of the mass lost through the outer Lagrangian point is also not simply determined. The reasonable assumption is that the mass lost from the outer Lagrangian point acquires the orbital velocity there. The outer Lagrangian mass loss with the wind velocities of up to around 600~\kmps, which are of the order of the orbital velocity, has been indeed seen in the P-Cygni profiles of stable nuclear burning WDs \citep{deufel1999olvel}. Therefore, we assume that the wind velocity in this phase is the same as the orbital velocity at the outer Lagrangian point. They are of the order of 100~\kmps.
It is also likely that the wind velocity decreases by the gravitational attraction. The earliest radio observation of SNe~Ia to constrain the CSM properties has been performed at 1~day after the explosion \citep{chomiuk2016ialimits}. The forward shock is at around $10^{14}~\mathrm{cm}$ at this epoch, while the outer Lagrangian point is at around $10^{11}~\mathrm{cm}$. In such a case, the wind velocity could be reduced by a factor of $\simeq 30$.

The mass-loss rates and wind velocities obtained in this way are summarized in Fig.~\ref{fig:allch}. The progenitor systems in this phase has relatively low mass-loss rates ($\sim 10^{-7}~\Msunpyr$) with the wind velocity of the order of $\sim 100~\kmps$ or less.

\subsection{Weak He-shell flash phase}\label{sec:weakhe}
The mass loss in the weak He-shell flash phase
is triggered by the He-burning shell flash at the surface of the WD. The flash is not strong enough to reduce the WD mass and it grows. A small amount of mass is blown from the system. The mass growth rate of a WD in this phase is estimated by using the prescription of \citet{kato2004mgrate} in the binary evolution model. The difference between the mass transfer rate and the mass growth rate is assumed to be ejected from the system as a wind. The mass-loss rates are found to be between $\simeq 3\times 10^{-9}~\Msunpyr$ and $\simeq 3\times 10^{-8}~\Msunpyr$ (Fig.~\ref{fig:allch}).

Because the weak He-shell flash is triggered by the nuclear burning at the surface of the accreting WD, we assume that the wind velocity becomes similar to those of novae. \citet{yaron2005nova} estimate that nova ejecta from a 1.4~\Msun\ WD have velocities of around 1000~\kmps\ when the mass accretion rate is $10^{-7}-10^{-6}~\Msunpyr$, which matches with our models. Therefore, we simply assume that the wind velocity is 1000~\kmps\ in this study.

The CSM properties estimated in this way are summarized in Fig.~\ref{fig:allch}. The systems in the weak He-shell flash phase have relatively low CSM density due to the small mass-loss rates and the large wind velocities.

Although we assume a smooth wind in constraining the CSM properties, the shell flash can actually form a CSM with many shells \citep[e.g.,][]{chomiuk2012sn11feradio}. Then, the CSM around SN~Ia progenitors in the weak He-shell flash phase likely has a less dense component than we find in Fig.~\ref{fig:allch} between the shells. Such a shell-like structure can also exist around SN~Ia progenitors with H-rich donor stars \citep{chomiuk2012sn11feradio}.

\subsection{Optically thick wind phase}\label{sec:optthick}
We assume that the transferred mass that exceeds the critical mass accretion rate is ejected from the system as an optically thick wind when a system is in the optically thick wind phase. The mass-loss rates of the systems in the optically thick wind phase mostly exceed $10^{-7}~\Msunpyr$ (Fig.~\ref{fig:allch}). With the high accretion rate and strong He-shell burning, the WD would form an extended He envelope reaching the Roche lobe. We assume that the mass loss occurs around the Roche radius and the wind velocity would be the escape velocity from the He envelope extended to the Roche lobe radius. Then, the wind velocity becomes around $1000~\kmps$ (Fig.~\ref{fig:allch}).

\begin{figure}
	\includegraphics[width=\columnwidth]{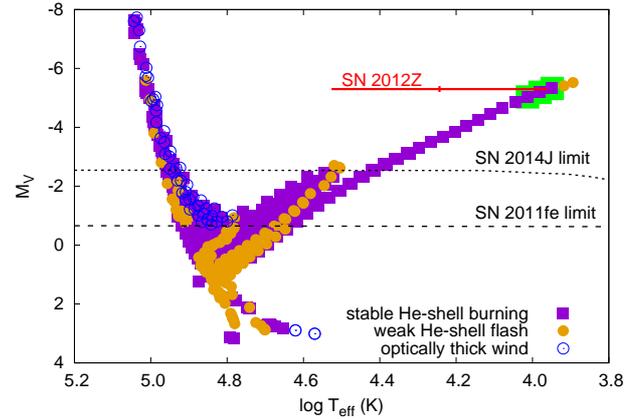}
    \caption{
    Effective temperature ($T_\mathrm{eff}$) and absolute $V$ band magnitude ($M_V$) of He star donors at the time of SN~Ia explosions. The effective temperature is obtained by the binary evolution calculation. $M_V$ is estimated by taking the bolometric luminosity obtained by the binary evolution model and applying the bolometric correction of \citet{torres2010bc}. The same method is applied by \citet{li2011sn11felimit}. The $M_V$ limits for SN~2011fe \citep{li2011sn11felimit} and SN~2014J \citep{kelly2014sn14Jlimit} are shown.
    The companion star of the SN~2012Z progenitor reported by \citet{mccully2014sn12Z} is shown and the models consistent with it are marked with green.
    }
    \label{fig:hrdbc}
\end{figure}

\begin{figure}
	\includegraphics[width=\columnwidth]{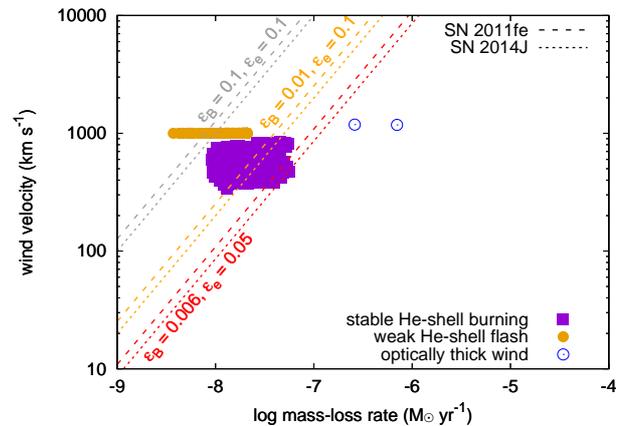}
    \caption{
    The same as Fig.~\ref{fig:allch}, but only for those with He donor luminosity below the limit of SN~2011fe in Fig.~\ref{fig:hrdbc}.
    }
    \label{fig:allch_11fe}
\end{figure}

\section{Discussion}\label{sec:discussion}
\subsection{Comparison with SN~2011fe and SN~2014J}\label{sec:11fe14J}
The expected CSM properties for the He star donor channel of SNe~Ia based on the binary evolution model of \citet{wang2009heprog} are summarized in Fig.~\ref{fig:allch}. The CSM properties around SNe~Ia are well constrained by radio observations of SNe~Ia (see \citealt{chomiuk2016ialimits} for a summary). No radio signals from SNe~Ia have been observed so far.
The deepest constraints on the CSM density around SNe~Ia come from SN~2011fe \citep{chomiuk2012sn11feradio,horesh2012sn11feradiox} and SN~2014J \citep{perez-torres2014sn14jradio,chomiuk2014sn14jradio,chandler2014sn14jradio}, which give the upper limits for the CSM density in Fig.~\ref{fig:allch}.
The right-side region of the lines are excluded by the radio observations.

Radio emission from SNe~Ia originates from synchrotron emission from relativistic electrons accelerated at the shock wave between the SN ejecta and CSM. However, the microphysics at a shock wave is quite uncertain. For example, a fraction of kinetic energy injected to the shock wave that is converted to magnetic field ($\varepsilon_B$) and a fraction that is used to the relativistic electron acceleration ($\varepsilon_e$) are still not well constrained, although they play a very important role in estimating radio emission from SNe~Ia. If we assume that the power-law index of the relativistic electron number density ($p$) is 3 and the outer density structure of SN~Ia ejecta is proportional to $r^{-7}$, the CSM density constraint becomes proportional to $(\varepsilon_B\varepsilon_e)^{-0.7}$ \citep[e.g.,][]{moriya2013lbvradio}.

The CSM density constraints with three different combinations of $\varepsilon_B$ and $\varepsilon_e$ are shown in Fig.~\ref{fig:allch}.
$\varepsilon_B$ and $\varepsilon_e$ are often assumed to be both 0.1, but several lines of arguments are against this assumption \citep[e.g.,][]{fransson1998sn93Jradio,bjornsson2004sn02ap,maeda2012shock,soderberg2012sn11dh,kamble2016sn13df,kundu2017iaradio}. In the previous studies of SN~Ia radio observations \citep[e.g.,][]{chomiuk2012sn11feradio,perez-torres2014sn14jradio,kundu2017iaradio}, the combination of $(\varepsilon_B, \varepsilon_e)=(0.01, 0.1)$ is also investigated. \citet{maeda2012shock} proposed lower values for both $\varepsilon_B$ and $\varepsilon_e$ by fitting the radio and X-ray observations of well-observed SN~IIb 2011dh and the best parameter estimate suggested is $(\varepsilon_B, \varepsilon_e)=(0.006, 0.05)$. We also adopted this combination in Fig.~\ref{fig:allch}. \citet{bjornsson2004sn02ap} and \citet{kamble2016sn13df} also suggest similarly small values for SN~Ic 2002ap and SN~IIb 2013df, respectively. Because the shock microphysics does not depend on SN types, this estimate should be applicable in SNe~Ia as well. 

The progenitor systems in the optically thick wind phase at the time of the explosion are mostly excluded in all the microphysics assumption as found in the previous studies \citep[e.g.,][]{chomiuk2012sn11feradio}. 
The progenitor systems at the stable He-shell burning phase are mostly excluded if we assume $(\varepsilon_B, \varepsilon_e)=(0.1, 0.1)$. However, this combination is not likely and we argue that the remaining two combinations of $\varepsilon_B$ and $\varepsilon_e$ provide more realistic constraints when we take the aforementioned parameter constraints from core-collapse SN observations into account. 
In this case, a large fraction of the systems at the stable He-shell burning are not excluded by the radio observations of SN~2011fe and SN~2014J. All the systems at the weak He-shell flash phase at the time of the explosion are not excluded under the probable microphysics assumptions.

\begin{figure*}
	\includegraphics[width=\columnwidth]{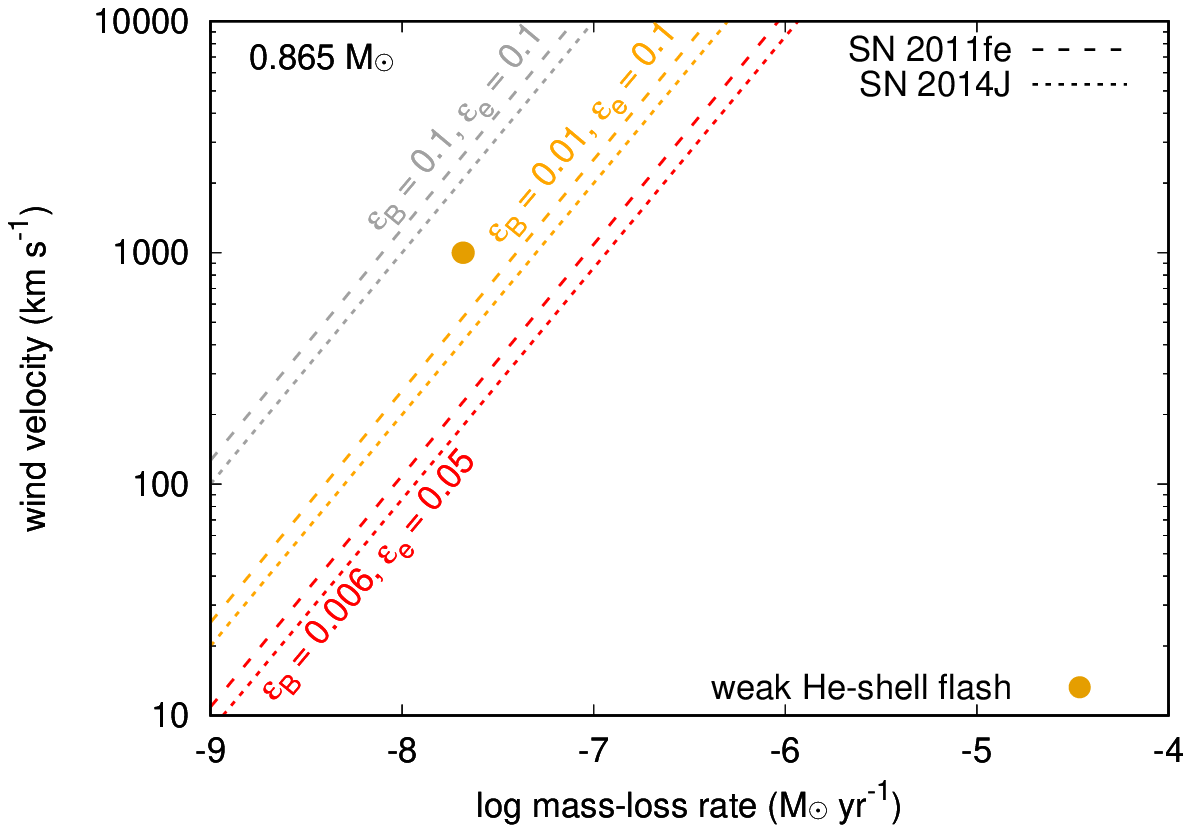}
	\includegraphics[width=\columnwidth]{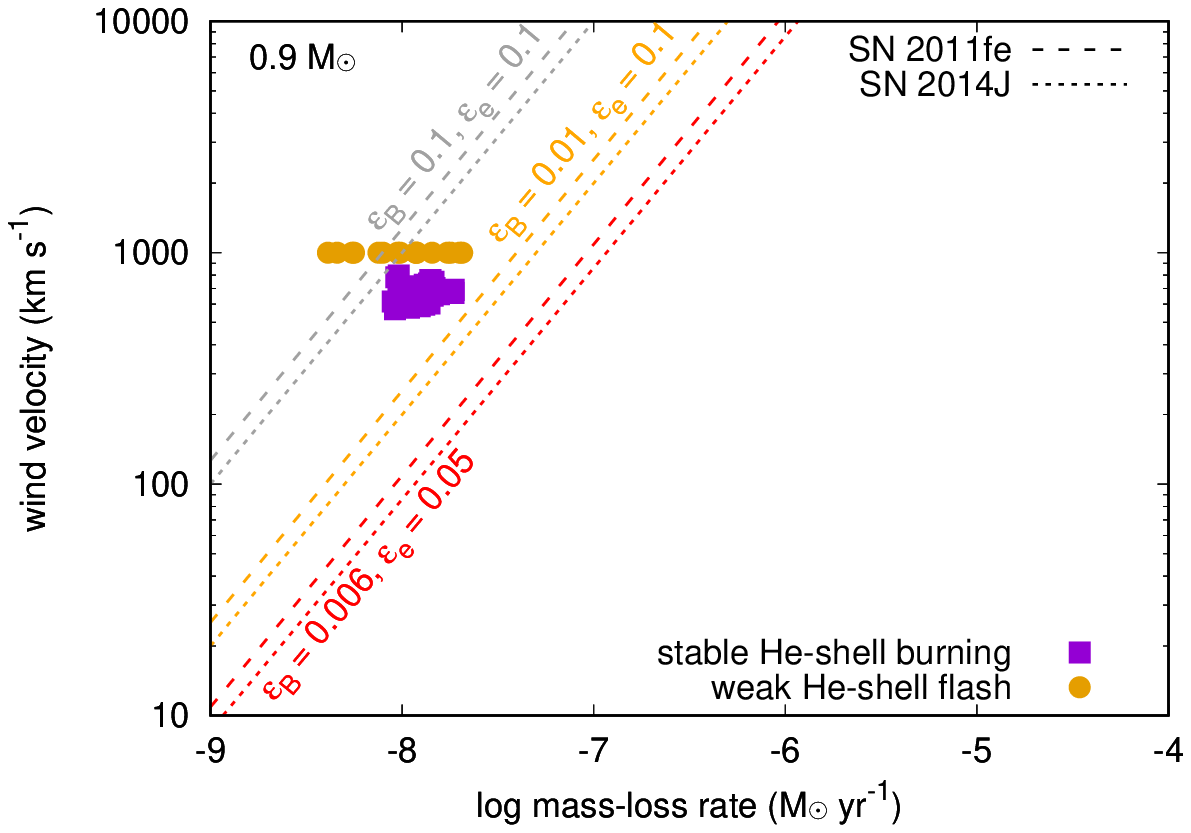} \\
	\includegraphics[width=\columnwidth]{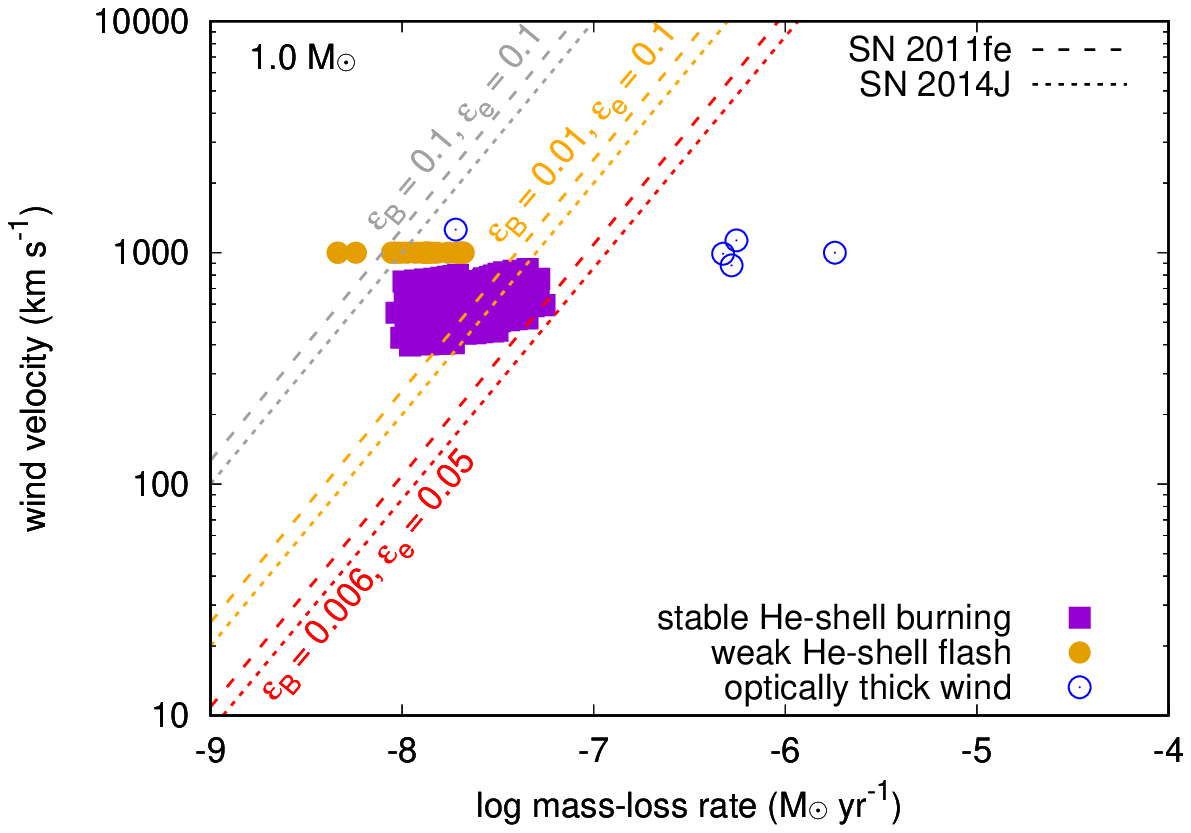}
	\includegraphics[width=\columnwidth]{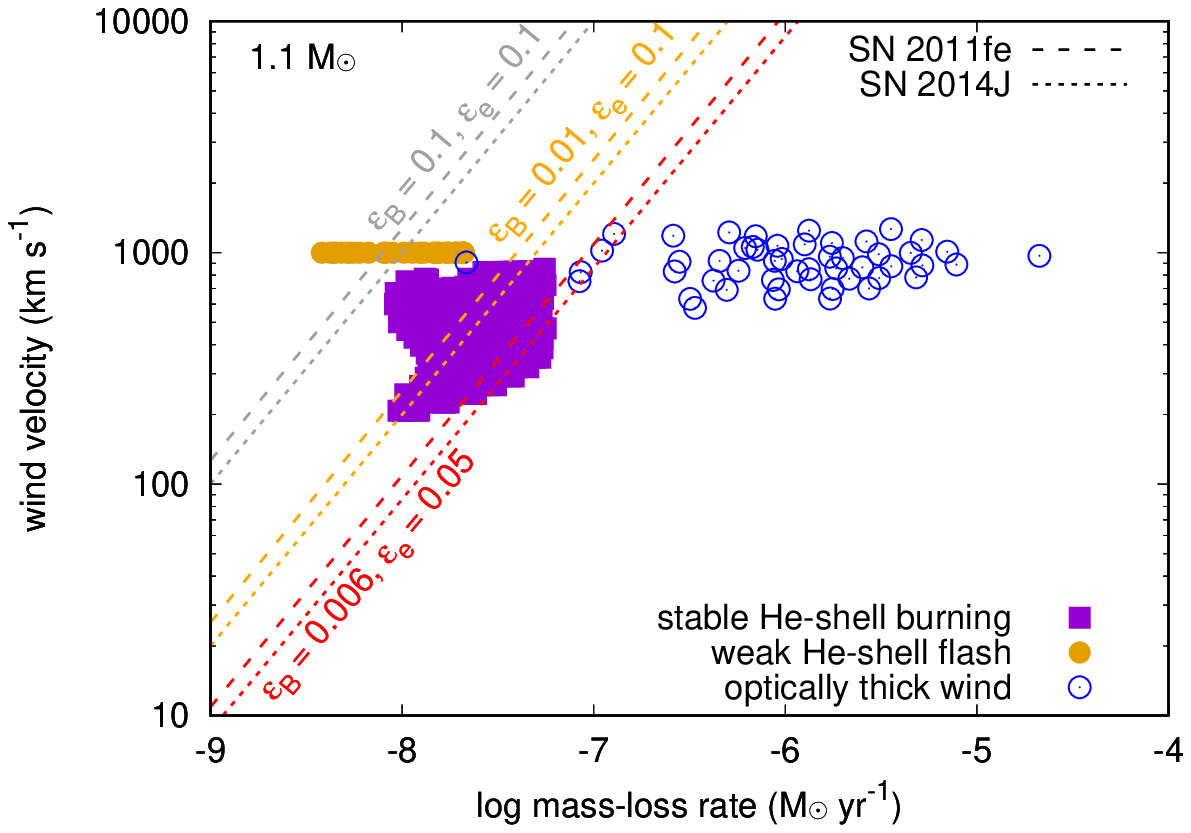} \\	
	\includegraphics[width=\columnwidth]{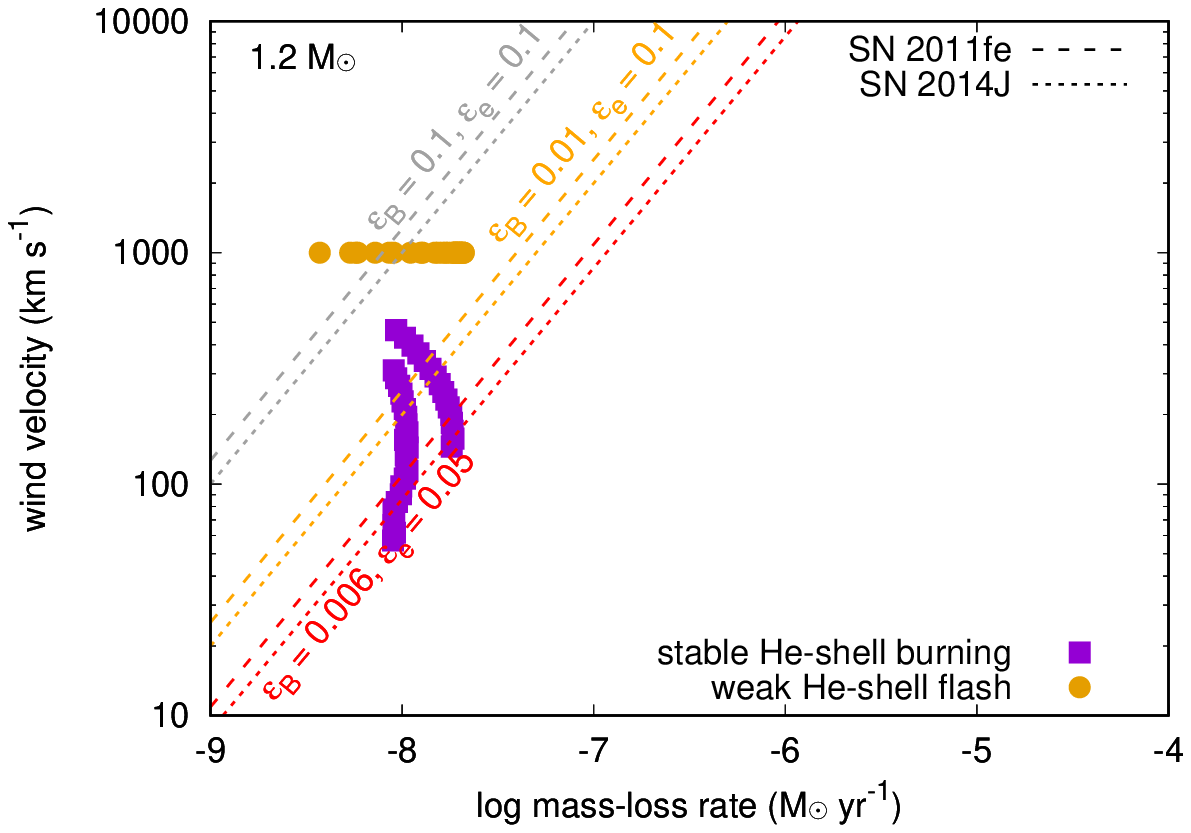}
    \caption{
    CSM properties of SNe~Ia from the He star donor channel with different initial WD masses. The initial WD masses are shown at the top right of each panel. See Fig.~\ref{fig:allch} for further details.
    }
    \label{fig:iniwd}
\end{figure*}

Population synthesis models generally predict that the prompt SNe~Ia are dominated by the He star donor channel \citep[e.g.,][]{wang2009hepop,ruiter2009iapop,claeys2014iapop,liu2015sdiax}. SN~2014J has been suggested to be a prompt SN~Ia \citep{nielsen2014sn14Jpromp} and it showed unexpectedly early gamma-ray emission from the \Ni\ decay which might be related to He accretion from its companion \citep{diehl2014sn14j56ni}. Although SN~2011fe did not have He star donor signatures, it appeared in a spiral arm of a star-forming galaxy \citep[e.g.,][]{li2011sn11felimit} so it might also be from the prompt SN~Ia population. Being prompt SNe~Ia, both SN~2011fe and SN~2014J may have had a He star donor.
In addition, the magnitude limits for the companion stars of SN~2011fe and SN~2014J obtained 
by the pre-SN images do not exclude a large fraction of He star donors predicted by the population synthesis model. 
Fig.~\ref{fig:allch_11fe} shows the CSM properties of SN~Ia progenitor systems that have the He donor stars below the upper limit of SN~2011fe in Fig.~\ref{fig:hrdbc}. We can see that most systems avoid the constraints from the radio observations.
Especially, a major fraction of SNe~Ia from the He star donor are predicted to be in the weak He-shell flash phase (Section~\ref{sec:rates}) and most of the progenitor systems in this phase are not excluded by the pre-SN images.
These systems originate from a wide variety of the initial WD mass, He star donor mass, and initial orbital period (cf. Fig.~\ref{fig:iniwd}).
In summary, the current radio observations and pre-explosion images leave many possible systems for the He star donor channel in both SN~2014J and SN~2011fe. We note that subdwarf B star companions are also suggested to avoid these observational constraints \citep{2019MNRAS.482.5651M}.

\begin{table}
	\centering
	\caption{Galactic SN~Ia rates from the He~star donor channel.}
	\label{tab:rate}
	\begin{tabular}{lcc} 
		\hline
		 & \multicolumn{2}{c}{rate ($10^{-3}~\mathrm{yr^{-1}}$)} \\
		 & $\alpha_\mathrm{CE}\lambda=0.5$ & $\alpha_\mathrm{CE}\lambda=1.5$\\
		\hline \hline
		total & 1.07 & 1.10 \\
		\hline
		stable He-shell burning & 0.25 (23\%) & 0.59 (54\%) \\
		weak He-shell flash     & 0.82 (77\%) & 0.51 (46\%) \\
		\hline
		\hline
		optically thick wind$^a$ & 0.11 & 0.11 \\
	\hline
	\multicolumn{3}{l}{$^a$ AIC progenitors.}
	\end{tabular}
\end{table}

\begin{table}
	\centering
	\caption{
	Galactic SN~Ia rates from the He~star donor channel having the He companion stars fainter than the SN~2011fe limit as in Fig.~\ref{fig:hrdbc}.
	}
	\label{tab:rate_11fe}
	\begin{tabular}{lcc} 
		\hline
		 & \multicolumn{2}{c}{rate ($10^{-3}~\mathrm{yr^{-1}}$)} \\
		 & $\alpha_\mathrm{CE}\lambda=0.5$ & $\alpha_\mathrm{CE}\lambda=1.5$\\
		\hline \hline
		total & 0.53 & 0.28 \\
		\hline
		stable He-shell burning & 0.15 (30\%) & 0.14 (54\%) \\
		weak He-shell flash     & 0.35 (70\%) & 0.13 (46\%) \\
		\hline
		\hline
		optically thick wind$^a$ & 0.03 & 0.01 \\
	\hline
	\multicolumn{3}{l}{$^a$ AIC progenitors.}
	\end{tabular}
\end{table}

\subsection{SN~2012Z}
A possible He companion star at the explosion site of SN~Iax 2012Z is discovered \citep{mccully2014sn12Z}. We show this candidate companion star property in Fig.~\ref{fig:hrdbc}. The effective temperature is estimated by its color by assuming the relation provided by \citet{torres2010bc}. We marked 3 models that are consistent with this possible companion star in Fig.~\ref{fig:hrdbc} as well as in Fig.~\ref{fig:allch}. They all have the initial He donor mass of 1.05~\Msun\ and their initial orbital periods are $1.8-2.0$~days. The initial WD mass is 1.2~\Msun. The final He donor mass is 0.86~\Msun\ with the final orbital periods of $1.9-2.1$~days. The WD accretion rate at the time of the explosion is $\simeq 10^{-6}~\Msunpyr$. These models are in the stable He-shell burning phase and have relatively high CSM density (Fig.~\ref{fig:allch}). However, \citet{liu2015heiax} found that such a progenitor system for SN~2012Z is rather rare. We refer to \citet{liu2015heiax} for the further investigation of SNe~Iax from the He-star donor channel.

\subsection{Rates}\label{sec:rates}
\citet{wang2009hepop} estimated Galactic SN~Ia rates from each He star donor progenitor system based on the binary evolution models of \citet{wang2009heprog}. Their results are summarized in Table~\ref{tab:rate}. \citet{wang2009hepop} adopted the rapid binary evolution code by \citet{hurley2000bps,hurley2002bps} for their population synthesis. When a WD + He star system is in the parameter range for SN~Ia progenitors obtained by \citet{wang2009heprog} at the onset of the Roche-lobe overflow, the system is assume to produce a SN~Ia. We assume that the systems at the optically thick wind phase when the WD reaches 1.378~\Msun\ undergo AIC instead of SNe~Ia. An important but uncertain parameter of the binary population synthesis in estimating the SN~Ia rates is $\alpha_\mathrm{CE}\lambda$, where $\alpha_\mathrm{CE}$ is the common-envelope ejection efficiency and $\lambda$ is the envelope binding energy parameter \citep{dewi2000alphalambda,tauris2001alphalambda}. The two parameters, i.e., 
$\alpha_\mathrm{CE}\lambda=0.5$ and 1.5 are adopted to estimate the SN Ia rate from the He star channel. 

The total Galactic SN~Ia rate from the He star donor channel is estimated to be $\simeq 1.1\times 10^{-3}~\mathrm{yr^{-1}}$ in both $\alpha_\mathrm{CE}\lambda=0.5$ and $\alpha_\mathrm{CE}\lambda=1.5$ cases. However, the fraction of the progenitor systems in each phase depends strongly on the assumed $\alpha_\mathrm{CE}\lambda$ as summarized in Table~\ref{tab:rate}. With $\alpha_\mathrm{CE}\lambda=0.5$, the progenitor system in the weak He-shell flash phase dominates SNe~Ia with He star donors, but both systems in the weak He-shell flash phase and the stable He-shell burning phase almost equally contribute to SNe~Ia when $\alpha_\mathrm{CE}\lambda = 1.5$. We note that $\alpha_\mathrm{CE}\lambda$ is constrained to be around $0.1-1$ in low- and intermediate-mass stars \citep[e.g.,][]{zorotovic2010alphalambda,demarco2011alphalambda,davis2012alphalambda} so the systems in the weak He-shell flash may be more common in SNe~Ia from He star donors.

We showed that many systems from the He star donor channel avoid the detection limit of the SN~2011fe progenitor system in Section~\ref{sec:11fe14J}. Table~\ref{tab:rate_11fe} shows the SN~Ia rates only for these systems avoiding the SN~2011fe limit. We find about a half of SNe~Ia from this channel have a He companion star fainter than the SN~2011fe limit in the case of $\alpha_\mathrm{CE}\lambda = 0.5$, while about 30\% of the systems avoid the limit in the case of $\alpha_\mathrm{CE}\lambda = 1.5$.

\subsection{Initial WD mass dependence}
A wide range of initial WD masses is adopted in the binary evolution model of \citet{wang2009heprog}. Fig.~\ref{fig:iniwd} shows the CSM properties with different initial WD masses. The majority of the progenitor systems have the initial WD mass of 1.1~\Msun\ regardless of the phases at the time of the explosions. No clear dependence of the CSM properties on the initial WD mass exists and it is difficult to constrain the initial WD mass through radio observations.

\subsection{Uncertainties}
We have adopted the binary stellar evolution model of \citet{wang2009heprog} in this study. However, it should be kept in mind that uncertainties in the stellar evolution calculations can affect the expected CSM properties presented in this work. For example, the WD mass growth rates under given accretion rates are still uncertain \citep[e.g.,][]{brooks2016heacc,wong2019schwab}. Different mass-retention efficiencies onto a WD are expected to affect the results of our binary calculations \citep[e.g.,][]{bours2013,ruiter2013}. In addition, rotation of WDs is ignored in the stellar evolution model we adopted, which can also change the WD evolution \citep{hachisu2012rotatingwd}. A spin-down process could be important to determine the moment of the SN explosions if rotation of WDs is considered. Depending on the spin-down timescale, the CSM around the progenitor system could diffuse and reach a density similar to that of the interstellar medium, leading to the lack of radio emission \citep{justham2011}. Observations of the radio emission from SNe~Ia with the He star donor channel would give us a better understanding of these uncertainties.

\subsection{Future prospects}
We have shown that the current radio observations are not deep enough to exclude most of the He star donor channel of the SD model. It will be necessary to conduct radio observations of nearby SNe~Ia like SN~2011fe and SN~2014J with deeper limits to detect radio emission from SN Ia from the He star donor channel. For this purpose, Five hundred meter Aperture Spherical radio Telescope (FAST) will shortly be a powerful tool. The upper limits of the radio emission from SN~2011fe and SN~2014J were $\sim 10~\mu\mathrm{Jy}$ \citep{chomiuk2016ialimits} at $\sim 1~\mathrm{GHz}$. FAST is expected to reach $\sim 1\mu\mathrm{Jy}$ at $\sim 1~\mathrm{GHz}$ in a few hours and can make the flux limit deeper by one order of magnitude. With this depth, a large fraction of SNe~Ia from the He star donor channel especially in the stable He-shell burning phase are predicted to be detectable (Fig.~\ref{fig:allch}). Square Kilometer Array (SKA) will eventually allow us to observe SNe~Ia down to $\sim 0.1~\mu\mathrm{Jy}$ \citep{perez-torres2015ask} and we should be able to make the final conclusion on the He star donor channel.

There are several other suggested ways to test the He star donor SD channel. For example, the SN ejecta are predicted to strip the surface of the He star donors and the SN ejecta will be contaminated by the stripped He \citep{pan2010stripping,pan2012stripping,liu2013hestripping}. The amount of He stripped by the SN ejecta is predicted to be $\sim 0.01~\Msun$ \citep{pan2010stripping,pan2012stripping,liu2013hestripping}. The SN ejecta contaminated by the stripped He are predicted to have He emission in the nebular phase \citep{botyanszki2018stheemi}. No He emission is observed from SN~2011fe and SN~2014J in the nebular phase and the stripped mass is constrained to be less than $\sim 0.01~\Msun$ \citep{tucker2019iastr}. SN~2012Z, which is suggested to have a He star companion from the pre-explosion image \citep{mccully2014sn12Z}, however, did not show the predicted He emission and the stripped mass is also constrained to be less than $\sim 0.01~\Msun$ \citep{tucker2019iastr}. The predicted masses and the upper limits are now comparable and it is still difficult to make a firm conclusion on the mass stripping with the current observations. Further late-phase observations of SNe~Ia are required.  

Another intriguing prediction is that the He star donor may get brighter in $\sim 10-100$~years after the explosion because of the heat provided by the SN ejecta collision \citep{pan2013hecompexp}. Especially for the case of SN~2011fe, the faint He star donor that was not detected by the pre-explosion image could be now bright enough to be detected \citep{pan2013hecompexp}. This prediction could also be tested by SN~2012Z with the He star companion \citep{mccully2014sn12Z}.

\section{Conclusions}\label{sec:conclusions}
We have investigated the CSM properties around SN~Ia progenitors from the SD model having a He star donor. We estimated the CSM properties based on the binary evolution calculations of WD + He star systems by \citet{wang2009heprog}. The binary evolution calculations suggest that SNe~Ia from the He star donor channel occur during the stable He-shell burning phase or the weak He-shell flash phase. We found that the current deepest radio observations from SN~2011fe and SN~2014J cannot exclude the possibilities of SN~Ia progenitors with He star donors in either phase (Fig.~\ref{fig:allch}). We also found that their pre-explosion images cannot exclude most of the progenitor system in the weak He-shell flash phase, which is predicted to dominate SNe~Ia with He star donors (Fig.~\ref{fig:hrdbc}). Therefore, both SN~2011fe and SN~2014J could both be prompt SNe~Ia. Future radio observations by FAST and SKA are likely to detect radio emission from the CSM interaction in SNe~Ia from the He star donor channel if a SN~Ia occurs as close as SN~2011fe or SN~2014J.



\section*{Acknowledgements}
We thank the referee for a constructive report. TJM thanks Sunny Wong for comments.
TJM is supported by the Grants-in-Aid for Scientific Research of the Japan Society for the Promotion of Science (JP17H02864 and JP18K13585).
BW is supported by the National Natural Science Foundation of China (Nos 11873085, 11673059 and 11521303), the Chinese Academy of Sciences (No QYZDB-SSW-SYS001), and the Yunnan Province (No 2018FB005).
ZWL is supported by the National Natural Science Foundation of China (NSFC, No. 11873016) and 100 Talents Programme of the Chinese Academy of Sciences.




\bibliographystyle{mnras}
\bibliography{mnras} 







\bsp	
\label{lastpage}
\end{document}